\documentclass[seceq]{ptptex}

\usepackage{graphicx}
\graphicspath{{fig/}}
\usepackage{wrapft}

\title{
Simulation of depositions of a Lennard-Jones cluster on a crystalline surface%
}

\author{
Kuniyasu \textsc{Saitoh}$^{1,}$\footnote{E-mail: saitoh@yukawa.kyoto-u.ac.jp} 
and Hisao \textsc{Hayakawa}$^{2,}$
}

\inst{
$^{1,}$ $^2$Yukawa Institute for Theoretical Physics, Kyoto University,\\ 
Kyoto 606-8502, Japan
}

\abst{
Depositions of amorphous Lennard-Jones clusters on a crystalline surface are numerically investigated. 
From the results of the molecular dynamics simulation, we found that the deposited clusters exhibit a transition from multilayered adsorption to monolayered adsorption at a critical incident speed.
Employing the energy conservation law, we can explain the behavior of the ratio of the number of atoms adsorbed on the substrate to the cluster size.
The boundary shape of the deposited cluster depends strongly on the incident speed and some unstable modes grow during the spread of the deposited cluster on the substrate.
We also discuss the wettability between different Lennard-Jones atoms.
}

\begin{document}

\maketitle

\section{Introduction}
A nanocluster containing 10 - 10,000 molecules exhibits intermediate properties between bulk materials and individual molecules. 
Recently, there has been growing interest in the physics of nanoclusters.\cite{book1,book2,book3,rev1,rev3} \ 
In particular, it is important to investigate depositions of nanoclusters on solid surfaces 
for the construction of high-quality films used in nanoscale electronic devices and photonic devices.\cite{rev2} \ 

The ionized cluster beam (ICB) technique was developed by Yamada \textit{et al}.\cite{icb1,icb2,icb3} \ 
The ICB technique is used to produce atomic clusters by employing adiabatic expansion of condensed vapour through a nozzle into a high vacuum region. 
In the ICB technique, clusters are ionized by electron impact and then accelerated toward a substrate.
Because the ICB technique controls the translational kinetic energy of the cluster, 
there have been many experimental and theoretical studies aimed at understanding the influence of the incident velocities of the cluster.\cite{incident1,incident2,incident3} \ 

The outcome of such a cluster impact is largely influenced  by the incident velocity, as seen from the phase diagrams in references.\cite{diag1,diag2} \ 
If the translational kinetic energy per atom becomes too large, the cluster can damage the substrate,
and the cluster can break into pieces after the impact.\cite{high1,high2,high3,high4,high5,high6,fragment1} \ 
However, if the translational kinetic energy per atom is less than $100$ eV, the cluster is adsorbed on the surface or reflected by the surface.
Awasthi \textit{et al}. carried out molecular dynamics simulations for collisions of Lennard-Jones clusters with weakly attractive surfaces.\cite{reflect1,reflect2} \ 
They discovered that the cluster rebounds when the translational kinetic energy of the cluster is larger than the adhesion energy.
Moreover, they clarified that a transition from adhesion to rebound occurs at the critical translational kinetic energy.
J\"{a}rvi \textit{et al}. carried out molecular dynamics simulations of low-energy deposition of individual metal clusters on a (100) surface.\cite{epitaxy1,epitaxy2} \ 
They revealed that the heat generated by the impact partially or completely melts the deposited cluster.
As a consequence, the atoms in the cluster are rearranged and adjusted to the atomic structure of the substrate.
They found the maximum size at which single clusters align epitaxially on the substrate.

Recently, Kuninaka and Hayakawa carried out molecular dynamics simulations of two identical colliding clusters 
and investigated impact phenomena of nanoclusters subject to thermal fluctuations.\cite{kuninaka1} \ 
They found super-rebound events in which the restitution coefficient is larger than $1$.
They confirmed the validity of macroscopic quasi-static theory of cohesive collisions.\cite{bri1} \ 
This suggests that the research of nanoclusters are relevant even for the study for fine powders whose diameters are ranged from $100$ nm to $1$ $\mu$ m.\cite{Tomas1,Castellanos} \ 
They also revealed the mechanism responsible for the super-rebound process, the normal rebound and the merging.

Although early numerical studies assumed that the clusters are highly crystallised, we also need to know the properties of amorphous clusters.
Indeed, it is easy to form metastable amorphous clusters in terms of the quench process from high temperature liquids.\cite{amorphous,freezing1,freezing2,freezing3,freezing4,freezing5} \ 

The main purpose of our paper is to understand the behavior of the deposited amorphous Lennard-Jones clusters on the crystalline surface at zero temperature. 
Here, we report on our molecular dynamics simulation of the depositions with the small incident energies per atom which lie in the so-called soft-landing regime ($0-2$ eV). 
In addition, we report on the wettability between different Lennard-Jones atoms.

From the analysis of the final configurations of the deposited clusters, 
we find the existence of a morphological phase transition from the hemi-spherical droplet to the monolayer film at the critical incident speed. 
The multilayered adsorption state is described by the energy conservation law.
Furthermore, we find that there are some unstable modes of the boundary shape of the deposited cluster.

The organization of this paper is as follows. 
\S 2 consists of two subsections.
\S 2.1 explains the model of our numerical simulation.
We explain our setup of cluster depositions in \S 2.2. 
\S 3 consists of five subsections.
\S 3.1 exhibits some time evolutions of impact processes.
\S 3.2 explains how the cluster size and the cluster adsorption parameter depend on the incident speed after the impact.
\S 3.3 discusses the transition from partial wetting to perfect wetting of deposited clusters.
\S 3.4 explains the morphological change of the final configuration of adsorbed atoms in clusters.
\S 3.5 discusses the wettability between different Lennard-Jones atoms.
In \S 4, we discuss our numerical results and we summarize the conclusion. 

\section{Molecular dynamics simulation \label{sec:MD}}

\subsection{Model}

\begin{figure}
\centerline{\includegraphics[width = 12 cm]{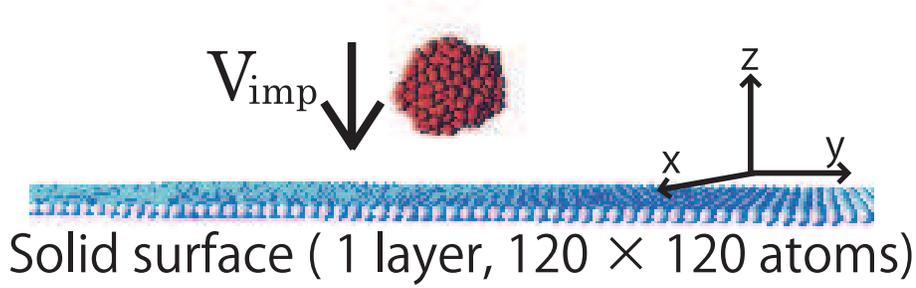}}%
\caption{ (Color online)
One snapshot of our simulation of a nanocluster deposition.
The incident cluster contains $300$ atoms which are bounded by the Lennard-Jones potential.
The substrate consists of a single layer ( $120 \times 120$ ) atoms on a triangular lattice.
\label{fig:setting}}
\end{figure}

In order to investigate the nanocluster depositions on a substrate, we perform a molecular dynamics simulation. 
Figure \ref{fig:setting} displays one snapshot of our numerical simulation.
Because we are interested in neutral nanoclusters and substrates, the electrostatic interaction between atoms is not considered.
We assume that the potential energy of interaction between two atoms can be described by the Lennard-Jones ( LJ ) potential: 
\begin{equation}
U(r_{ij}) = 4\epsilon_{\alpha\beta} \biggl\{ \left( \frac{\sigma_{\alpha\beta}}{r_{ij}} \right)^{12} - \left( \frac{\sigma_{\alpha\beta}}{r_{ij}} \right)^6 \biggr\} ,
\label{eq:LJ}
\end{equation}
where subscripts $\alpha$ and $\beta$ specify the species of LJ atoms, $r_{ij}$ is the distance between two atoms labeled by $i$ and $j$.
Here, $\epsilon_{\alpha\beta}$ and $\sigma_{\alpha\beta}$ are respectively the strength of the interaction and the diameter of the repulsive core between $\alpha$ atom and $\beta$ atom.
If $\alpha \neq \beta$, we adopt the cross parameters of the LJ potential by using the Lorentz-Berthelot rule as
\begin{equation}
\sigma_{\rm \alpha\beta} = \frac{(\sigma_{\alpha}+\sigma_{\beta})}{2},  \quad
\epsilon_{\rm \alpha\beta} = \sqrt{\epsilon_{\alpha}\epsilon_{\beta}}.
\label{eq:L-B}
\end{equation}
We mainly investigate the case that the cluster and the substrate are constructed by same atoms, A.
Here, we borrow the values of LJ parameters and mass of a typical inert gas, argon.
Therefore, $\epsilon_{\rm{AA}}$, $\sigma_{\rm{AA}}$ and the mass of an A atom $m_{\rm{A}}$ are $1.65 \times 10^{-21} \rm{J}$, $3.405 \mathrm{\mathring{A}}$ and $6.63 \times 10^{-26} \rm{kg}$,
respectively.\cite{book6,ar-c1,ar-c2} \
On the other hand, to study the influence of the interaction energy between different atoms of the cluster and the substrate, 
we introduce a C atom as the constituent of the substrate in \S 3.5.
We use the values of LJ parameters and mass of carbon to the C atom, in which $\epsilon_{\rm{CC}}$, $\sigma_{\rm{CC}}$ and 
the mass of C atom $m_{\rm{C}}$ are $3.86 \times 10^{-22} \rm{J}$, $3.354 \mathrm{\mathring{A}}$ and $1.99 \times 10^{-26} \rm{kg}$, respectively.\cite{book6,ar-c1,ar-c2} \
We should note that the interaction energy between the cluster and the substrate, $\epsilon_{\rm{AC}}$, is several times weaker than $\epsilon_{\rm{AA}}$.
In the following, we omit the subscripts of the LJ parameters of the interaction between the A atoms and we adopt simplified notations $\epsilon$, $\sigma$.
We also adopt $m$ as the mass of an A atom. We use $\epsilon$, $\sigma$ and $m$ as the units of energy, length and mass, respectively.
Thus, the unit time is given by $\tau=\sqrt{m\sigma^2/\epsilon}$.

We use a single layer surface which involves $120 \times 120$ atoms on a triangular lattice as the substrate with the periodic boundary condition.\cite{diff1,diff2}\ 
We set the lattice constant to $2^{1/6}\sigma_{\rm{\alpha\alpha}}$ ( $\alpha = \rm{A}, \rm{C}$ ) as the equilibrium distance between atoms.
To avoid the destruction of the substrate, each atom of the substrate is also tethered to its equilibrium position by an elastic spring.
In actual impacts of nanoclusters on substrates,
 the energy induced by an impact is relaxed to the internal motion of the atoms of the bulk region of the substrate.
To represent such a energy relaxation process, we simply introduce the viscous force proportional to its velocity.
The introduction of the viscous force has another advantage to reduce the unrealistic boundary effects.
Indeed, if we do not introduce the viscous force, the acoustic wave would be transmitted across the boundary.
The atom of the substrate at $\mathbf{r}_i$ satisfies the equation of motion
\begin{equation}
m_{\alpha} \frac{d^2 \mathbf{r}_i}{dt^2} = - \sum_j \frac{d}{d\mathbf{r}_i} U(r_{ij}) - k (\mathbf{r}_i - \mathbf{r}_i^{eq}) - \lambda \frac{d \mathbf{r}_i}{dt},
\label{eq:subeq}
\end{equation}
where $\sum_j$ is a summation over the interacting pairs $i$ and $j$, and $\mathbf{r}_i^{eq}$ is its equilibrium position, and $m_{\alpha}$ ( $\alpha = \rm{A}, \rm{C}$ ) is the mass of an atom of the substrate.
Because we consider the substrate as a rigid surface, we use the spring constant $k=1.0\times 10^3 \epsilon/\sigma^2$.
We simply adopt the value $\lambda=1.0 \sqrt{m \epsilon}/\sigma$ for the coefficient of viscosity.

We adopt the velocity Verlet method for numerical integration of the equation of motion for each atom with the time step $dt=1.0 \times 10^{-3} \sqrt{m \sigma^2/\epsilon}$.
To reduce computational costs, we introduce the cut-off length $\sigma_{\rm cut}=3.0\sigma$ to the LJ potential,
and we adopt the periodic boundary conditions in the horizontal $xy$ directions and the free boundary condition in the vertical $z$ direction.
It should be noted that, the viscous force is evaluated as the value at the previous time step for the numerical integration of Eq.~(\ref{eq:subeq}).

\subsection{Setup}
\begin{figure}
\centerline{\includegraphics[width = 12 cm]{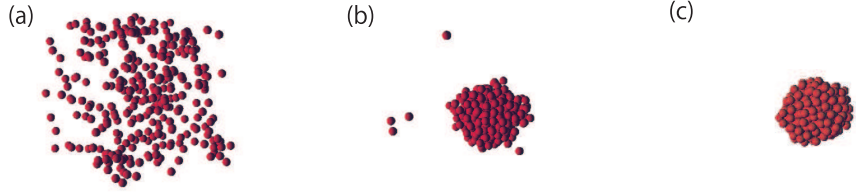}}%
\caption{ (Color online)
Illustration of a LJ cluster formation consisting of $300$ atoms formation by temperature quench method.
(a) The initial configuration of atoms in the gas  phase at $T=1.0 \epsilon$. (b) A liquid cluster obtained from the quench into $T=0.5 \epsilon$. (c) An amorphous cluster obtained from the quench into $T=0.01 \epsilon$.
\label{fig:quechcluster}}
\end{figure}

We make a LJ cluster by the temperature quench \cite{quench1} into the metastable phase of LJ fluid.\cite{lj-diag1} \ 
We prepare $32, 108, 255, 300, 500$ and $862$ atoms in a periodic box and equilibrate at the temperature $T=1.0 \epsilon$ with the number density $0.05 \sigma^{-3}$ in the gas state ( Fig.~\ref{fig:quechcluster} (a) ).
It should be noted that the unit of the temperature becomes $\epsilon$, because we set the Boltzmann constant to be unity.
To equilibrate the gas at a specific temperature, we adopt the velocity scaling method and perform until $\tau=2000 \sigma \sqrt{m/\epsilon}$ for the relaxation to a local equilibrium state. 
We have confirmed the equilibration of the total energy in the initial relaxation process, and we quench the gas to $T=0.5 \epsilon$.
After an equilibration, a weakly bounded liquid cluster is formed ( Fig.~\ref{fig:quechcluster} (b) ) and is quenched to $T=0.01 \epsilon$ to make it rigid. 
This two step quench is adopted to form one cluster from an initial gas state.
Indeed, if we quench the system into $T=0.01 \epsilon$ directly, many small clusters appear. 
After this equilibration process, we obtain an amorphous cluster ( Fig.~\ref{fig:quechcluster} (c) ).
We place the amorphous cluster at $10\sigma$ above the substrate and give the cluster the translational velocity $V_{{\rm imp}}$ to make it collide against the substrate.
It should be noted that the amorphous cluster is metastable to keep its shape within our observation time.
The incident angle of the cluster to the substrate normal is zero. 
The incident speed of the cluster is ranged from $V_{{\rm imp}}=0.1\sqrt{\epsilon/m}$ to $V_{{\rm imp}}=5.0\sqrt{\epsilon/m}$.

\section{Results}
\subsection{Time evolution of impact processes \label{sec:result1}}
\begin{figure}
\centerline{\includegraphics[width = 12 cm]{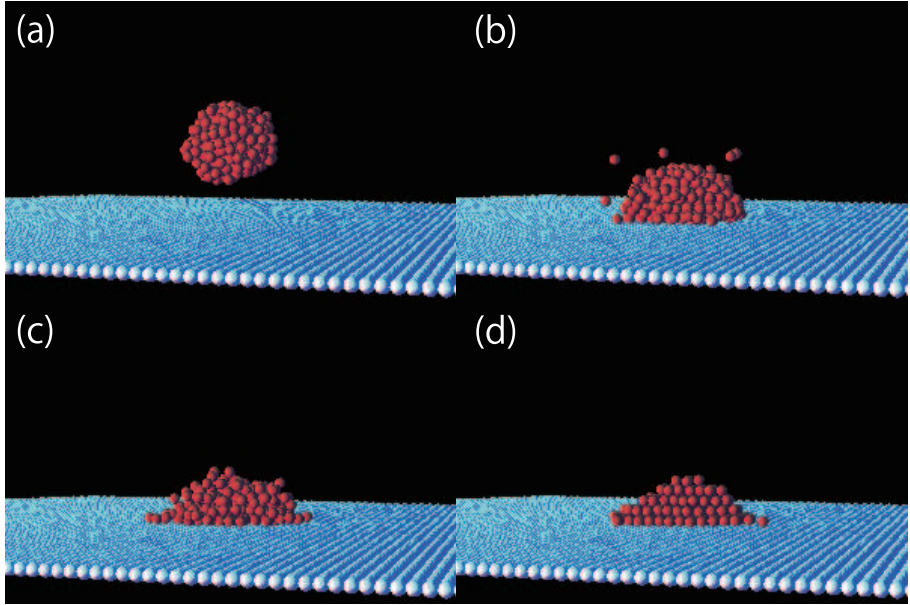}}%
\caption{ (Color online)
The time evolution of an impact process of a LJ cluster of $300$ atoms on the substrate, where the incident speed is $V_{{\rm imp}}=2.0\sqrt{\epsilon/m}$.
See the text in details.
\label{fig:impact1}}
\end{figure}
\begin{figure}
\centerline{\includegraphics[width = 12 cm]{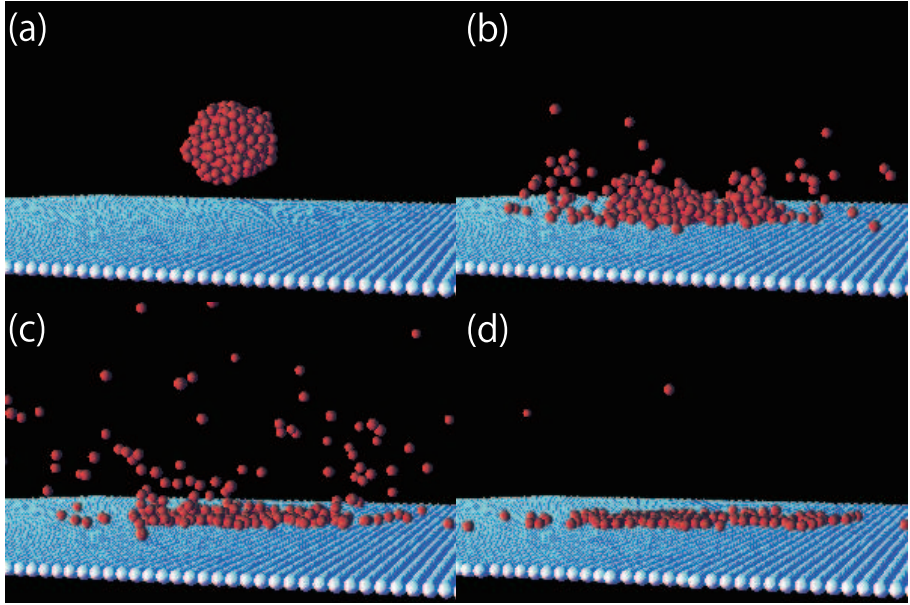}}%
\caption{ (Color online)
The time evolution of an impact process of a LJ cluster of $300$ atoms on the substrate, where the incident speed is $V_{{\rm imp}}=4.0\sqrt{\epsilon/m}$.
See the text in details.
\label{fig:impact2}}
\end{figure}
Figures \ref{fig:impact1} and \ref{fig:impact2} display the time evolutions of the impact process of the LJ cluster of $300$ atoms on the crystalline surface.
Figures~\ref{fig:impact1} (a)-(d) represent the case of $V_{{\rm imp}}=2.0\sqrt{\epsilon/m}$, and Figs.~\ref{fig:impact2} (a)-(d) are the case of $V_{{\rm imp}}=4.0\sqrt{\epsilon/m}$.

The incident cluster moves toward the substrate with its translational speed $V_{{\rm imp}}$ ( Figs.~\ref{fig:impact1} (a) and \ref{fig:impact2} (a) ), 
and hits the substrate ( Figs.~\ref{fig:impact1} (b) and \ref{fig:impact2} (b) ).
After the hitting, the cluster is only deformed to be a hemi-sphere ( Fig.~\ref{fig:impact1} (c) ) for the small incident speed.  
If the incident speed is, however,  larger than a critical value, the deposited cluster is split into many pieces ( Fig.~\ref{fig:impact2} (c) ).
After time goes on, the deposited cluster is adsorbed on the substrate and settles into the final configuration ( Figs.~\ref{fig:impact1} (d) and \ref{fig:impact2} (d) ).

We observe that the impact process and the final configuration depend strongly on the incident speed $V_{{\rm imp}}$.
In the case of $V_{{\rm imp}} < 1.7\sqrt{\epsilon/m}$, no atoms can escape from the cluster during the impact.
By contrast, some atoms evaporate during the impact process for $V_{{\rm imp}} \geq 1.7\sqrt{\epsilon/m}$.
If the incident speed is relatively small, the final configuration is a hemi-sphere on the substrate, as in the case of a partial wetting of a liquid droplet on a dry surface.
The deformation is larger   as the incident speed increases.
Above $V_{{\rm imp}}=3.3\sqrt{\epsilon/m}$, the deposited cluster is completely split into fragments and the absorbed atoms on the substrate form a monolayer coverage.
Above $V_{{\rm imp}} = 4.5\sqrt{\epsilon/m}$, the deposited cluster is burst into fragments, and the absorbed coverage is no longer characterized by one cluster.

At the moment of the impact, the temperature of the deposited cluster increases because the initial kinetic energy is transformed into internal motion.
Then, the temperature decreases due to the heat conduction into the bulk region of the material 
through the contact area.\cite{epitaxy1,epitaxy2,dissipation1} \ 
The configuration of the deposited cluster is changed into an energetically favorite position during this cooling process.
Furthermore, the atomic structure of the deposited cluster is adjusted to the substrate.
Finally, the configuration is frozen because of the loss of temperature. 

\subsection{The incident speed dependencies of the scaled cluster size and the cluster adsorption parameter \label{sec:result2}}
\begin{figure}
\centerline{\includegraphics[width = 14 cm]{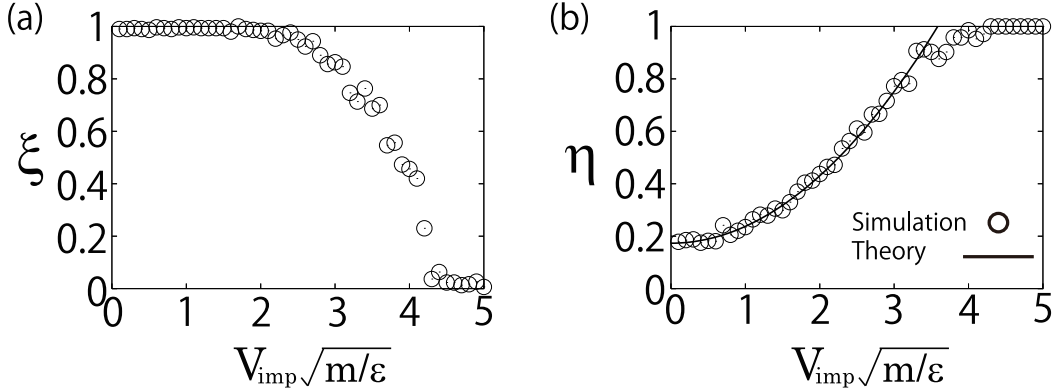}}%
\caption{
(a) A plot of the scaled cluster size $\xi$  and (b) a plot of the cluster adsorption parameter $\eta$  as the functions of the incident speed for the cluster of $300$ atoms.
\label{fig:adsorption}}
\end{figure}
In our simulation, the main cluster is detected by using the clustering algorithm.\cite{book4} \
Following the Allen and Tildesley, we adopt the critical atom separation $r_C=1.6\sigma$.
After the cluster settles into the final configuration, we represent $N_{cls}$ as the number of atoms in the cluster. 
With the aid of the number of atoms in the cluster before the impact $N$, we introduce the scaled cluster size :
\begin{equation}
\xi \equiv \frac{N_{cls}}{N}
\label{eq:xi}
\end{equation}
If $\xi=1$, no atoms can escape from the cluster after the impact. 
On the other hand, if $\xi<1$, evaporation of some atoms exists during the cluster impact. 

Let us define an absorbed atom in the cluster if
an atom at $\mathbf{r}$ in the cluster  satisfies the relation $|\mathbf{r}-\mathbf{r}_s| < r_C$,
where $\mathbf{r}_s$ is the position of its nearest neighbor constituent of the substrate.
Using the number of these adsorbed atoms $N_{adh}$, we can introduce the cluster adsorption parameter :
\begin{equation}
\eta \equiv 
\frac{N_{adh}}{N_{cls}}
\label{eq:eta}
\end{equation}
If $\eta<1$, the cluster is regarded as a multilayered adsorption.
However, if $\eta=1$, the deposited cluster is perfectly spread on the substrate, and it is a monolayered adsorption.

\begin{figure}
\centerline{\includegraphics[width = 14 cm]{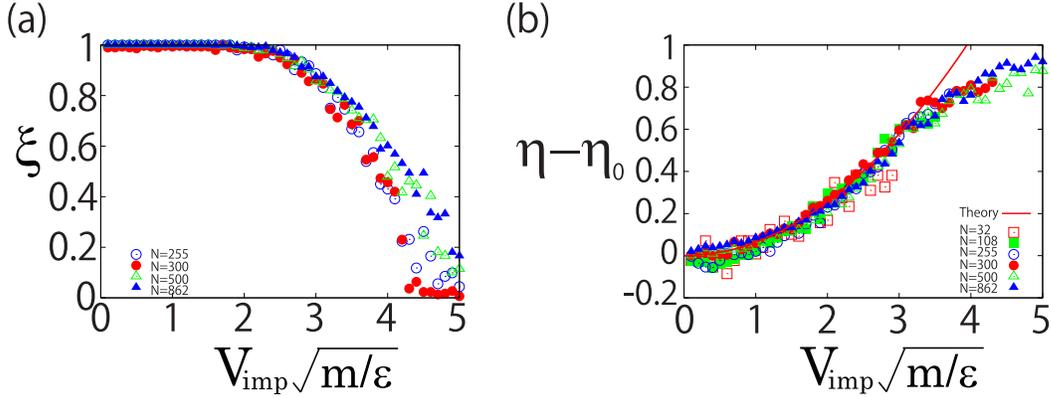}}%
\caption{ (Color online)
(a)  Plot of the scaled cluster size $\xi$, and (b) plot of the scaled cluster adsorption parameter $\eta$ as the functions of the incident speed.
\label{fig:N_adsorption}}
\end{figure}

Figures \ref{fig:adsorption} (a) and (b) plot the incident speed dependences of $\xi$ and $\eta$ for the cluster of $300$ atoms.
We find that $\xi$ equals to 1 below $V_{{\rm imp}}=1.7\sqrt{\epsilon/m}$, but
it decreases above $V_{{\rm imp}}=1.7\sqrt{\epsilon/m}$.
On the other hand, $\eta$ increases with the incident velocity below $V_{{\rm imp}}=3.3\sqrt{\epsilon/m}$,
 but it is satisfied to $\eta \simeq 1$ above $V_{{\rm imp}}=3.3\sqrt{\epsilon/m}$.
Figure \ref{fig:N_adsorption} (a) plots several results on $\xi$ for $N = 255, 300, 500$ and $862$.
While Fig.~\ref{fig:N_adsorption} (b) is  $\eta-\eta_0$  for $N = 32, 108, 255, 300, 500$ and $862$, where $\eta_0$ is $\eta$ at $V_{{\rm imp}}=0$.
It seems that $\eta-\eta_0$ is independent of the size of clusters, while $\xi$ exhibits weak size dependence.

How can we understand the behaviors in Figs.~\ref{fig:adsorption} and \ref{fig:N_adsorption}?
During the impact, the temperature in the cluster increases because the kinetic energy is transformed into internal motion.\cite{epitaxy1} \ 
We assume that the energy flux to the substrate $\Phi_{bulk}$ during the impact is small and 
the temperature becomes maximum $T_{max}$ when the speed of the center of mass of the cluster becomes zero.
Thus, the energy conservation law can be written as
\begin{equation}
\frac{1}{2}mNV_{{\rm imp}}^2+\frac{3}{2}NT_0 \simeq \frac{3}{2}NT_{max} + \Delta S,
\label{eq:heatup}
\end{equation}
where $T_0$ is the temperature of the cluster before the impact, and $\Delta S$ is the change of the surface energy.
With the introduction of the surface tension $\gamma$, the height of the deposited cluster $h$, the contact radius of the deposited cluster $R$ and the ratio $\phi=h/R$,
 $\Delta S$ is given by
\begin{equation}
\Delta S = \gamma(2\pi\phi R^2 - 4\pi R_0^2),
\label{eq:surface}
\end{equation}
where $R_0$ is the mean radius of the cluster before the impact.
Introducing the mean area fraction of the contact area $\rho_{adh} = N_{adh}/\pi R^2$, $\Delta S$ can be rewritten as
\begin{equation}
\Delta S = \gamma \left( \frac{2N_{adh}}{\rho_{adh}}-4\pi R_0^2 \right) .
\label{eq:surface2}
\end{equation}

From Eq.~(\ref{eq:heatup}), $T_{max}$ satisfies
\begin{equation}
T_{max} = T_0 + \frac{m}{3}V_{{\rm imp}}^2 - \frac{2\Delta S}{3N},
\label{eq:tmax}
\end{equation}
Because the binding energy per atom in the cluster is roughly equal to $\epsilon$, the evaporation takes place at $T_{max} \simeq \epsilon$. 
In our simulation, $T_0$ is much smaller than $T_{max}$, and the last term on the right hand side of Eq.~(\ref{eq:tmax}) is negligible for large $N$. 
Thus, the evaporation is believed to take place near $V_{{\rm imp}} \simeq \sqrt{3\epsilon/m}$. 
In Figs.~\ref{fig:adsorption} (a) and \ref{fig:N_adsorption} (a), the scaled cluster size becomes $\xi < 1$ 
above $V_{{\rm imp}} = 1.7\sqrt{\epsilon/m}$, which is consistent with the above estimation.
For the clusters with 32 and 108 atoms, $\xi$ decreases faster than the other cases.
In such cases, we cannot ignore the last term on the right hand side of Eq.~(\ref{eq:tmax}).

During the impact, an evaporated atom carries away the volume energy $u_V$ which is the potential energy per atom and the kinetic energy $\frac{3}{2}T_{max}$ from the cluster. 
We assume that the internal energy of the deposited cluster decreases because of the energy flux to the bulk of the substrate.
Therefore, after the cluster settles into the final configuration, the energy conservation law can be written as
\begin{equation}
\frac{1}{2}mNV_{{\rm imp}}^2+\frac{3}{2}NT_0 = \Delta S + \Phi_{bulk} + (1-\xi)N\bar{E},
\label{eq:dissipate}
\end{equation}
where $\bar{E} = \frac{3}{2}T_{max}-u_V$ is the energy carried away by an evaporated atom.
Here, $(1-\xi)N$ represents the number of evaporated atoms.
If the incident kinetic energy is not large, the number of atoms in the cluster is approximately preserved during the impact.
Therefore, it is reasonable that the scaled cluster size satisfies $\xi \simeq 1$. 
If we assume $\phi \simeq 1$, the energy conservation law Eq.~(\ref{eq:dissipate}) can be simplified as
\begin{equation}
\eta = \frac{m\rho_{adh}}{4\gamma}V_{imp}^2 + \frac{\rho_{adh}}{2\gamma} \left( \frac{4\pi \gamma R_0^2}{N} - \frac{\Phi_{bulk}}{N} + \frac{3}{2}T_0 \right),
\label{eq:small_Vimp-eta}
\end{equation}
where we have used Eqs.~(\ref{eq:eta}) and (\ref{eq:surface2}).

We use $\rho_{adh} = 0.91$, because the adsorbed atoms should match the hexagonal lattice on the substrate.
The mean radius $R_0$ of a cluster consisting $N$ atoms satisfies $R_0=r_0 N^{1/3}$ 
where we use 
$r_0=0.68 \sigma$ as a fitting parameter.
The solid line in Fig.~\ref{fig:adsorption} (b) is the theoretical prediction ( Eq.~(\ref{eq:small_Vimp-eta}) ), 
where the surface tension $\gamma \simeq 3.5\epsilon/\sigma^2$ and 
the energy flux per atom $\varphi_{bulk} = \Phi_{bulk}/N \simeq 1.7\epsilon$ are other two fitting parameters.

The second term on the right hand side of Eq.~(\ref{eq:small_Vimp-eta}) can be written as
\begin{equation}
\eta_0 (N) = \frac{\rho_{adh}}{2\gamma} \left( 4\pi r_0^2 \gamma N^{-1/3} - \varphi_{bulk} + \frac{3}{2}T_0 \right).
\label{eq:eta0}
\end{equation}
It is interesting that $\eta-\eta_0 (N)$ is independent of the cluster size.
Figure \ref{fig:N_adsorption} (b) plots our numerical results $\eta-\eta_0 (N)$ for $\eta<1$, which support the validity of
the theoretical prediction ( the solid line ).

\subsection{Transition from partial wetting to perfect wetting of the deposited cluster \label{sec:result3}}
\begin{figure}
\centerline{\includegraphics[width = 14 cm]{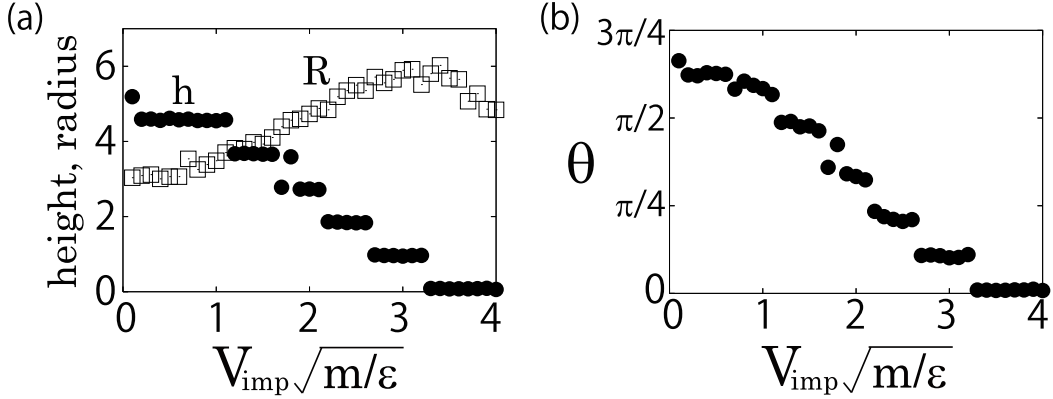}}%
\caption{
(a) Plots of the height $h$ ( filled circle ) and the contact radius $R$ ( open square ) and (b) a plot of the contact angle $\theta$  as the functions of the incident speed for the LJ cluster of $300$ atoms.
\label{fig:wetting}}
\end{figure}
Let us introduce the radius of the equimolar dividing surface ( Gibbs Surface ) \cite{book4}
\begin{equation}
R^2 = -\frac{1}{\rho_{adh}}\int_0^\infty \frac{d\rho(r)}{dr}r^2 dr
\label{eq:Gibbs}
\end{equation}
as the contact radius of a deposited cluster, 
where $\rho(r)$ is the area fraction of the contact area with radial distance from the center of mass of the adsorbed atoms in the cluster. 
We also define the cluster height $h$ as $z_{max}-z_0$, where $z_{max}$ is the maximum vertical position in the atoms in the cluster, 
and $z_0$ is the minimum vertical position. 
Assuming a meniscus shape to the deposited cluster, we geometrically calculate the contact angle $\theta$. 

Figure \ref{fig:wetting} displays $h$, $R$ and $\theta$ for the deposited cluster consisting of $300$ atoms as the functions of the incident speed. 
We observe that the cluster height $h$ decreases and the contact radius $R$ increases as the incident speed increases. 
Above $V_{{\rm imp}} = 3.3\sqrt{\epsilon/m}$, the height and the contact angle becomes zero, 
which implies that the deposited cluster becomes a monolayer film and is perfectly wetting on the substrate.
In this regime, the monolayer film is spread further and its boundary is partially chipped. Therefore the contact radius decreases.
The clusters consisting of $108,255,500$ and $864$ atoms are also perfectly wetting on the substrate at critical incident velocities.
However, it should be noted that the transition from multilayer film to monolayer film is only the morphological change of the deposited cluster.
In the case of the cluster consisting of $32$ atoms, the number of adsorbed atoms is too few to define the wetting parameters $h$, $R$ and $\theta$.

\subsection{Morphology of the final configuration of the adsorbed atoms in the cluster \label{sec:result4}}
\begin{figure}
\centerline{\includegraphics[width = 14 cm]{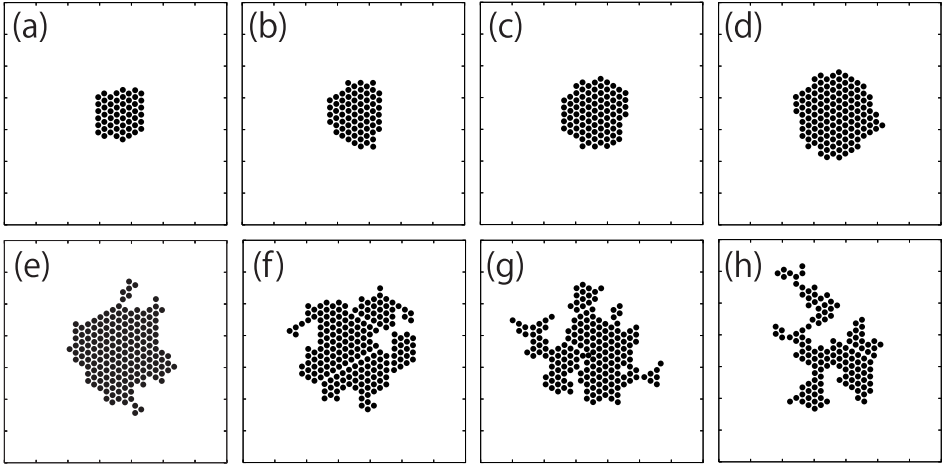}}%
\caption{ 
Configurations of the adsorbed atoms in a deposited cluster of $300$ atoms for each incident speed.
$V_{{\rm imp}}$ equals (a) $0.5 \sqrt{\epsilon/m}$, (b) $1.0 \sqrt{\epsilon/m}$, (c) $1.5 \sqrt{\epsilon/m}$, 
(d) $2.0 \sqrt{\epsilon/m}$, (e) $2.5 \sqrt{\epsilon/m}$, (f) $3.0 \sqrt{\epsilon/m}$, 
(g) $3.5 \sqrt{\epsilon/m}$, and (h) $4.0 \sqrt{\epsilon/m}$, respectively.
\label{fig:monolayer}}
\end{figure}
The boundary shape of the contact area depends strongly on the incident speed.
In order to investigate the morphology of the boundary shape, we define the radial distance of the boundary $r=f(\psi)$.
Here, $r$ and $\psi$ are the usual radial and azimuthal coordinates.
We take the origin to the center of mass of the adsorbed atoms in the cluster.
Moreover, we define a dimensionless variable $g(\psi)$ for the boundary \cite{boundary1} as
\begin{equation}
g(\psi) = \frac{f(\psi)-R}{R}.
\label{eq:mono_g}
\end{equation}
We also use its Fourier representations $g(\psi) = \sum_n g_n e^{in\psi} $, with the integer $n=0,\pm1,\pm2,\cdots$.

Figures \ref{fig:monolayer} (a)-(h) are the variety of the final horizontal configurations of the adsorbed atoms in the cluster consisting of $300$ atoms.
We find that there are three phases in the boundary shape.
Below $V_{{\rm imp}}=1.5\sqrt{\epsilon/m}$, the boundary shape is grainy ( Figs.~\ref{fig:monolayer} (a), (b) ), and $|g_n|^2$ has some peaks at higher modes.
This may be caused by the small number of adsorbed atoms.
In the case of $1.5\sqrt{\epsilon/m} < V_{{\rm imp}} < 3.0\sqrt{\epsilon/m}$, 
the deposited cluster is uniformly spread on the substrate ( Figs.~\ref{fig:monolayer} (c), (d), (e), (f) ).
Thus, peaks of $|g_n|^2$ vanish and their boundaries can be fitted by circles.
In the case of $3.0\sqrt{\epsilon/m} < V_{{\rm imp}} < 4.0\sqrt{\epsilon/m}$, the deposited cluster becomes a thin film or a monolayer film. 
In this regime, the boundary is partially chipped ( Figs.~\ref{fig:monolayer} (g), (h) ), where $|g_n|^2$ has intense peaks at some modes in our simulation.
Above $V_{{\rm imp}} = 4.0\sqrt{\epsilon/m}$, the deposited cluster is burst into fragments and the number of adsorbed atoms is too few to define $R$.
Thus, we can not define $g(\psi)$ in this regime.

The thermal fluctuation of a circular geometry step is estimated as $ \langle |g_n|^2 \rangle = T/2\pi\beta Rn^2$ from the equipartition of energy among the $g_n$.\cite{boundary1,book5} \ 
Here, $\beta$ is the step edge stiffness. 
In our simulation, the thermal fluctuation is estimated as $ \langle |g_n|^2 \rangle \sim 0.1$, 
while the $|g_n|^2$ has peaks ranging from $1000$ to $3000$ above $V_{{\rm imp}}=3.0\sqrt{\epsilon/m}$. 
Therefore this intensive peaks reflect on the growth of some unstable modes of $g_n$ during the spread of the deposited cluster on the substrate.

\subsection{Wettability between different Lennard-Jones atoms \label{sec:result5}}

\begin{figure}
\centerline{\includegraphics[width = 14 cm]{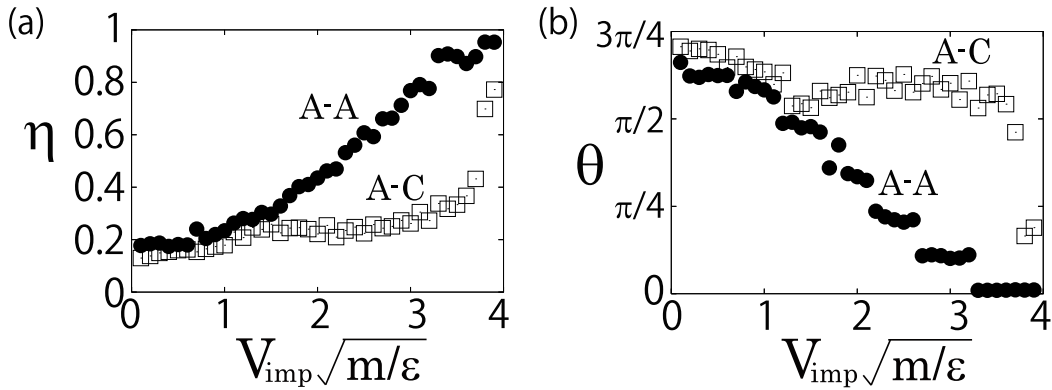}}%
\caption{
(a) Plots of the cluster adsorption parameter $\eta$, and 
(b) plots of the contact angle $\theta$ as the functions of the incident speed for both A-C case ( open square ) and A-A case ( filled circle ).
\label{fig:ar-c}}
\end{figure}
To investigate the influence of the interaction energy, we also perform the molecular dynamics simulation in which
the LJ parameters of A and C are used for the atoms in the cluster and the atoms of the substrate, respectively.
Henceforth, we call this situation A-C case.
The Lorentz-Berthelot rule Eq.~(\ref{eq:L-B}) is adopted to calculate the cross LJ parameters $\epsilon_{{\rm AC}}$ and $\sigma_{{\rm AC}}$.
Other simulation settings are the same as \S~\ref{sec:MD}. 

Figure \ref{fig:ar-c} displays the cluster adsorption parameter $\eta$ and the contact angle $\theta$ as the functions of the incident speed for the LJ cluster of 300 atoms.
We also plot the data in the case of the LJ parameters of A-A case.
We observe that the cluster adsorption parameter $\eta$ remains low value even when the impact speed is high ( Fig.~\ref{fig:ar-c} (a) ).
We also stress that any A cluster does not become a monolayer film in which $\theta$ becomes zero during the deposition onto a C surface.
This is because the wetted state of the argon cluster is unfavorite on the crystalline carbon surface,  which is resulted from $\epsilon_{{\rm AC}}<\epsilon$.
Thus, it is clear that not only the incident velocity, but also the choice of composites is important to determine the final configuration of the deposited cluster.

\section{Discussion and Conclusion}
In this paper, the incident kinetic energy per atom in the cluster is less than $2$ eV.
In this case, the damage of the substrate due to the impact of a cluster can be ignored and we considered a single layer substrate.
However, the influence of the interaction between the deposited cluster and the bulk of the substrate should be important.
In general, the adsorption state is strongly influenced by the surface temperature,\cite{surftemp1} \ but the substrate was assumed
to be at $T=0$ before the cluster impact in our simulation. Therefore, the influence of the surface temperature is also important for future study.
Moreover, we performed simulation only for one deposition event of the cluster at each incident speed.
Thus we should take ensemble average of impact processes for future study.

In conclusion, we found that deposited LJ clusters consisting $32$, $108$, $255$, $300$, $500$ and $862$ 
atoms exhibit a transition from multilayered adsorption to monolayered adsorption at the critical incident speed, $V_{\rm{imp}}=3.3\sqrt{\epsilon/m}$.
From our simulation, we clarified that the deposited clusters are perfectly wetting on the substrate above the critical incident speed.
Employing the energy conservation law, we estimated the critical value of the incident speed at which the evaporation begin to occur during the impact.
The estimated critical value, $V_{\rm{imp}}=1.7\sqrt{\epsilon/m}$, exhibits good agreement with our simulation results of $\xi$.
By using the energy conservation law, we also found that the scaled cluster adsorption parameter is independent of the cluster size 
and is proportional to $V_{\rm{imp}}^2$. These results exhibit good agreement with our simulation results.
We performed the Fourier analysis of $g(\psi)$ and found that some modes becomes unstable for $3.0\sqrt{\epsilon/m}<V_{\rm{imp}}<4.0\sqrt{\epsilon/m}$.
We also performed the molecular dynamics simulation of A-C case and we found that the A cluster does not become a monolayer film on the C surface.
Thus, we concluded that not only the incident speed, but also the strength of the interaction between the cluster and the substrate is important 
to form a monolayer film on a substrate.

\section*{Acknowledgements}
We would like to thank H. Kuninaka and H. Wada for their valuable comments. 
We would like to thank G. Paquette for his critical reading of this manuscript.
Parts of numerical computation in this work were carried out in computers of Yukawa Institute for Theoretical Physics, Kyoto University. 
This work was supported by the Global COE Program "The Next Generation of Physics, Spun from Universality and Emergence" from the Ministry of Education, Culture, Sports, Science and Technology ( MEXT ) of Japan. 
This work was also supported by the Research Fellowship of the Japan Society for the Promotion of Science for Young Scientists ( JSPS ), 
and the Grant-in-Aid of MEXT ( Grant Nos. 21015016 and 21540384 ).


%

\end{document}